\documentclass[preprint,proceedings]{rmaa}

\usepackage{paralist}

\SetYear{2006}
\SetConfTitle{First Light Science with the Gran Telescopio Canarias}

\title{THE CALIBRATION OF THE O/H ABUNDANCE INDICATORS FOR EXTRAGALACTIC 
\ion{H}{2} REGIONS BASED ON  \ion{O}{2} RECOMBINATION LINES}

\author{
  M. Peimbert\altaffilmark{1}, 
  A. Peimbert\altaffilmark{1}, C. Esteban\altaffilmark{2},
J. Garc\'{\i}a-Rojas\altaffilmark{2}, F. Bresolin\altaffilmark{3}, 
L. Carigi\altaffilmark{1}, M. T. Ruiz\altaffilmark{4}, 
and  A.~R.~L\'opez-S\'anchez\altaffilmark{2}}

\altaffiltext{1}{Instituto de Astronom\'\i{}a, U.N.A.M.
  M\'exico, D.F., M\'exico.}
\altaffiltext{2}{Instituto de Astrof\'\i{}sica de Canarias, 
Tenerife Espa\~na.}
\altaffiltext{3}{Institute for Astronomy, University of Hawaii,
Honolulu, Hawaii, U.S.A.}
\altaffiltext{4}{Departamento de Astronom\'{\i}a, Universidad de Chile, Chile.}
\shortauthor{Peimbert et al.}

\shorttitle{O/H extragalactic abundances}

\fulladdresses{
\item M. Peimbert:  
  Universidad Nacional Aut\'o\-no\-ma de M\'exico, 
  Apartado Postal 70-264, 04510
  M\'exico, D.F., M\'exico. (\email{peimbert@astroscu.unam.mx})}

\listofauthors{M. Peimbert, A. Peimbert, C. Esteban, J.
Garc\'{\i}a-Rojas, F. Bresolin, L. Carigi, M. T. Ruiz, and
A. R. L\'opez-S\'anchez }

\indexauthor{Peimbert, M.}
\indexauthor{Peimbert, A.}
\indexauthor{Esteban, C.}
\indexauthor{Garc\'{\i}a-Rojas, J.}
\indexauthor{Bresolin, F.}
\indexauthor{Carigi, L.}
\indexauthor{Ruiz, M. T.}
\indexauthor{L\'opez-S\'anchez, A. R.}

\abstract{Based on  \ion{O}{2} recombination lines we present a new 
calibration (called \ion{O}{2}$_{RL}$) of Pagel's $O_{23}$ indicator to 
determine the O/H abundance ratio in extragalactic 
\ion{H}{2} regions and emission line galaxies. The \ion{O}{2}$_{RL}$ calibration 
produces O/H abundances about a factor of two higher than those 
derived from the $T$(4363) method with $t^2$ = 0.00. The 
\ion{O}{2}$_{RL}$ calibration has 
implications for the study of different properties of 
emission line galaxies such as their metallicity, star formation 
rate, and initial mass function. The \ion{O}{2}$_{RL}$ calibration also 
affects the abundance determinations based on other O/H indicators, that include
collisionally excited lines, like those
known as $O_3N_2$, $N_2$, $S_{23}$, $Ar_3O_3$, and $S_3O_3$. We argue that 
the controversy between the $T$(4363) method and the photoionization 
models method to derive O/H values is mainly due to
temperature variations inside the observed \ion{H}{2} regions.}

\resumen{Presentamos una nueva calibraci\'on del indicador 
$O_{23}$ de Pagel para determinar los cocientes de O/H en
regiones \ion{H}{2} extragal\'acticas y galaxias con l\'i{}neas de emisi\'on.
Esta calibraci\'on la llamamos \ion{O}{2}$_{RL}$ y esta basada en l\'i{}neas
de recombinaci\'on de \ion{O}{2}.
Nuestra calibraci\'on produce abundancias de O/H alrededor de un 
factor de dos mayores que las obtenidas a partir del m\'etodo
$T_e$(4363) con $t^2$ = 0.00. La calibraci\'on \ion{O}{2}$_{RL}$ tiene implicaciones para el
estudio de diferentes propiedades de las galaxias con l\'i{}neas 
de emisi\'on tales como la metalicidad, la tasa de formaci\'on estelar,
y la funci\'on inicial de masas. La calibraci\'on \ion{O}{2}$_{RL}$
tambi\'en 
afecta aquellas determinaciones de abundancias basadas en otros
indicadores de O/H, que incluyen l\'i{}neas excitadas colisionalmente,
tales como los llamados 
$O_3N_2$, $N_2$, $S_{23}$, $Ar_3O_3$, y $S_3O_3$. Argumentamos que
la controversia entre el m\'etodo $T$(4363) y el m\'etodo basado en
modelos de fotoionizaci\'on para determinar O/H se debe principalmente
a la presencia de variaciones de temperatura dentro de las regiones
\ion{H}{2} observadas.}

\addkeyword{Galaxies: Abundances}
\addkeyword{H~II regions: Abundances}

\begin{document}
\maketitle

\section{Introduction}
\label{sec:intro}

The advent of large telescopes is permitting to observe \ion{H}{2}
regions in galaxies many tens of megaparsecs away from us and
emission line galaxies up to distances of $z\sim 3$. But since the amount of 
photons that we obtain from faraway objects is small, often
we have reliable information only for a few bright emission 
lines which has led to the idea of using different metallicity
indicators based on at most a handful of bright emission lines.

The most popular metallicity indicator was introduced by Pagel et al.
(1979, see also Edmunds \& Pagel 1984) and is indistinctly known as Pagel's, or $R_{23}$, or $O_{23}$
indicator, where $O_{23}\equiv 
I([$\ion{O}{2}$]\lambda 3727 + [$\ion{O}{3}$]\lambda\lambda 4959, 5007)/ I({\rm H}\beta)$. 
The $O_{23}$ indicator has been calibrated
with the O/H values based on two different methods: by using photoionization
models, that we will call $PIM$ calibrations or $PIM$ method,
and by using observational determinations of the O/H abundances 
based on the electron temperature derived
from the $I(4363)/I(5007)$ [$\ion{O}{3}$] ratio together with  the
$I(3727)/I({\rm H}\beta)$ and the $I(5007)/I({\rm H}\beta)$ line ratios, the so called 
$T(4363)$ method.

There are significant differences between the calibrations of Pagel's indicator
based on models (e. g. McCall et al. 1985;
Dopita \& Evans 1986; McGaugh 1991; Kobulnicky \& Kewley 2004) and the calibrations based on
observations and $T(4363)$ (e. g. Torres-Peimbert,
Peimbert, \& Fierro 1989; 
Pilyugin 2000, 2003; Castellanos, D\'{\i}az, \& Terlevich 2002; Pilyugin, \& Thuan 2005). The
differences in the O/H values are in the $0.2$ - $0.6$ dex range and could
be due mainly to the presence of temperature inhomogeneities over the observed
volume (e. g. Campbell 1988; Torres-Peimbert et al. 1989; McGaugh 1991;
Roy et al. 1996; Luridiana et al. 1999; Kobulnicky, Kennicutt, \& Pizagno 1999:
Kobulnicky, \& Kewley 2004). These differences need to be sorted out if we 
want to obtain absolute accuracies in 
O/H of the order of 0.1 dex or better. We will call these differences the
calibration controversy.

In this paper we present a qualitatively different calibration
of the $O_{23}$ indicator that is based on the intensity ratio of \ion{O}{2} 
recombination lines to  \ion{H}{1}  recombination lines, that we will call 
the \ion{O}{2}$_{RL}$ method. As of now the \ion{O}{2}$_{RL}$ calibration has been 
established only for objects in the high metallicity branch 
of the $O_{23}$ versus O/H relation and for log $O_{23}$ $>$ 0.5.

Preliminary discussions of the \ion{O}{2}$_{RL}$ calibration were presented by
Peimbert \& Peimbert (2003, 2005).

\section{The $O_{23}$ indicator}
\label{sec:OM}

The oxygen abundance by unit mass is an excellent tracer of the heavy 
element content $Z$ of a given \ion{H}{2} region because for extremely 
poor objects O constitutes about 60\% of the heavy elements by mass, while
for the present value of the ISM of the solar vicinity it amounts to
43\%. In the Local Group galaxies for a metallicity range of
0.00319 $< Z <$ 0.01990 (that corresponds to 8.15 $<$ 12 + log O/H $<$ 8.86), 
the fraction of $Z$ due to O  varies
from 53\% to 41\%, mainly due to the increase of N and C relative to
O as $Z$ increases (Peimbert 2003).

There are three different types of methods to calibrate the $O_{23}$
indicator: a) the $PIM$ method, b) the $T$(4363)
method, and c) the \ion{O}{2}$_{RL}$ method. We will discuss these three methods
and the causes for the O/H differences among them. In particular 
we will address the calibration controversy: why is it that 
the calibrations based on the $T$(4363) method yield abundances
from 0.2 to 0.4 dex lower than the calibrations based on photoionization 
models in the $O_{23}$ high metallicity branch for log $O_{23}$ $>$ 0.5.

\subsection{Calibration based on  \ion{O}{2} recombination lines}

Peimbert, Storey, \& Torres-Peimbert (1993) were the first 
to determine O/H values for gaseous nebulae based on the recombination
coefficients for \ion{O}{2} lines computed by Storey (1994). The temperature dependence
of the \ion{O}{2} lines is relatively weak and very similar to that of the
\ion{H}{1} lines, therefore the O$^{++}$/H$^+$ ratios are independent of the
electron temperature. Alternatively the O$^{++}$/H$^+$ ratios derived
from collisionally excited lines do depend strongly on the average
temperature, $T_0$, and the mean temperature square, $t^2$ (e. g.:
Peimbert 1967, Peimbert \& Costero 1969, Ruiz et al. 2003, Peimbert et al. 2004).

In \ion{H}{2} regions the recombination lines typically yield abundances
higher than the optical collisionally excited lines by factors in the 
2 to 3 range, if a value of
$t^2$ = 0.00 is adopted. This difference is due to the presence of strong temperature
variations that yield $t^2$ values in the 0.02 to 0.06 range. The $t^2$
determinations have been obtained by three different methods:
a) by comparing the temperature derived from the intensity ratio of 
the Balmer continuum to a recombination Balmer line with $T$(4363)
(Peimbert 1967) , b)
by determining $T_0$ and  $t^2$ from a least squares method using the line 
intensities of a large number of He I lines (Peimbert, Peimbert, \& Ruiz
2000, Peimbert, Peimbert, \& Luridiana 2002), and c) by computing the 
$t^2$ value needed to derive the same $N$(O$^{++}$)/$N$(H) ratio from
 \ion{O}{2} recombination lines and [O III] collisionally excited lines
(Peimbert, Storey, \& Torres-Peimbert 1993). For the best observed 
objects the three methods yield the same result, within the errors,  
supporting the presence of large temperature variations (e. g. Peimbert 
2003; Peimbert et al. 2004, 2005; Esteban et al. 2005). The presence of 
temperature variations affect strongly the $T$(4363) method, weakly
the $PIM$ method, and leave the \ion{O}{2}$_{RL}$
method unaffected, or in other words the \ion{O}{2}$_{RL}$
method is independent of the temperature structure
of the nebula.

In Table 1 we present the O/H values derived from the \ion{O}{2}$_{RL}$
method and the $T$(4363) method for  \ion{H}{2}  regions of nearby galaxies
and the Galaxy. The $N$(O$^{++}$)/$N$(H$^+$) values were derived from the
\ion{O}{2} recombination lines. Most of the $N$(O$^+$)/$N$(H$^+$) values were
derived from $I$(3727) together with  
$T$(5755/6584) and the $t^2$ value determined from several
methods, while the rest were derived from \ion{O}{1} recombination lines. In the first column we list the object, in the second column
the O/H value based on the \ion{O}{2}$_{RL}$ method, in the third column the
O/H value based on the $T$(4363) method, in the fourth column the
log $O_{23}$ observed value, and in the fifth column the  
ionization parameter $P$ defined as 
$P\equiv
I([$\ion{O}{3}$]\lambda\lambda 4959, 5007)/ 
I([$\ion{O}{2}$]\lambda 3727 + [$\ion{O}{3}$]\lambda\lambda 4959, 5007)$
(Pilyugin 2001). The average fraction of oxygen twice ionized 
in the sample presented in Table 1 amounts to 68\%.

\begin{table}[!t]\centering
  \setlength{\tabnotewidth}{\columnwidth}
  \tablecols{6}
  \caption{O/H Values derived by the \ion{O}{2}$_{RL}$ and the $T$(4363)
methods}
\label{tab:O/H}
{\small
\begin{tabular}{lc@{ \null }c@{ \null }c@{ \null }c@{ \null }c}
    \hline
    \hline
     & {\small O/H} & {\small O/H} & & & \\
{\small Object} & {\small O~II$_{RL}$} & {\small $T$(4363)} & {\small log
    $O_{23}$}  & {\small $P$} & {\small Ref.} \\
    \hline
M16        & 8.81 & 8.56 & 0.58 & 0.27 & 1  \\
M8         & 8.71 & 8.51 & 0.53 & 0.38 & 2  \\
M17        & 8.76 & 8.52 & 0.73 & 0.83 & 2  \\
M17        & 8.87 & 8.56 & 0.75 & 0.84 & 3  \\
M20        & 8.71 & 8.53 & 0.60 & 0.20 & 1  \\
NGC 3576   & 8.86 & 8.56 & 0.78 & 0.78 & 4  \\
NGC 3576   & 8.73 & 8.52 & 0.79 & 0.79 & 3  \\
Orion      & 8.71 & 8.51 & 0.77 & 0.86 & 5  \\
Orion      & 8.61 & 8.47 & 0.74 & 0.84 & 6  \\
NGC 3603   & 8.72 & 8.46 & 0.89 & 0.92 & 1  \\
S 311      & 8.57 & 8.39 & 0.72 & 0.32 & 7  \\
NGC 5461   & 8.81 & 8.56 & 0.80 & 0.74 & 8  \\
N11B(LMC)  & 8.74 & 8.41 & 0.80 & 0.70 & 3  \\
NGC 604    & 8.66 & 8.49 & 0.67 & 0.70 & 8  \\
30 Doradus & 8.57 & 8.34 & 0.89 & 0.86 & 3  \\
30 Doradus & 8.54 & 8.33 & 0.90 & 0.85 & 9  \\
N66 (SMC)  & 8.47 & 8.11 & 0.90 & 0.85 & 3  \\
NGC 5253   & 8.39 & 8.18 & 0.97 & 0.85 & 10 \\
NGC 6822   & 8.37 & 8.08 & 0.90 & 0.88 & 11 \\
NGC 2363   & 8.20 & 7.87 & 1.00 & 0.97 & 8  \\
    \hline
    \hline
    \tabnotetext{}{Given in 12 + log O/H.}
    \tabnotetext{}{1- Garc\'\i{}a-Rojas et al. (2006a);
2- Garc\'\i{}a-Rojas et al. (2006b); 3- Tsamis et al. (2003);
4- Garc\'\i{}a-Rojas et al.(2004); 5- Esteban et al. (2004);
6- Esteban et al. (1998); 7- Garc\'\i{}a-Rojas et al. (2005);
8- Esteban et al. (2002); 9- Peimbert (2003);
10- L\'opez-S\'anchez et al. (2006, zones A and B); 
11- Peimbert et al. (2005).}\end{tabular}
}
\end{table}

In Figure 1 we present the data of Table 1 and the calibration of Pagel's
indicator provided by the \ion{O}{2}$_{RL}$ method and the $T$(4363)
method.
Pilyugin (2001) has found that the 
$O_{23}$ indicator depends strongly on the ionization parameter
$P$
and that for a given $O_{23}$ value the higher the $P$ 
value the higher the O/H value. The amount of objects with measured
\ion{O}{2} recombination line intensities or accurate $t^2$ values is very small
and it is not possible to produce an absolute calibration for different
$P$ values, but of the 20 objects in Table 1 sixteen present 0.70 $< P <$ 0.97,
and based on them we have produced a calibration for $P$ = 0.8. In these
calibrations we have also
made use of the four objects with 0.20 $< P <$ 0.38 including the relative increase in O/H predicted by the $T$(4363) method for a change from $P$ = 0.3
to $P$ = 0.8, that amounts to $\sim$ 0.2 dex (Pilyugin \& Thuan 2005).

\begin{figure}[!t]
  \includegraphics[width=\columnwidth]{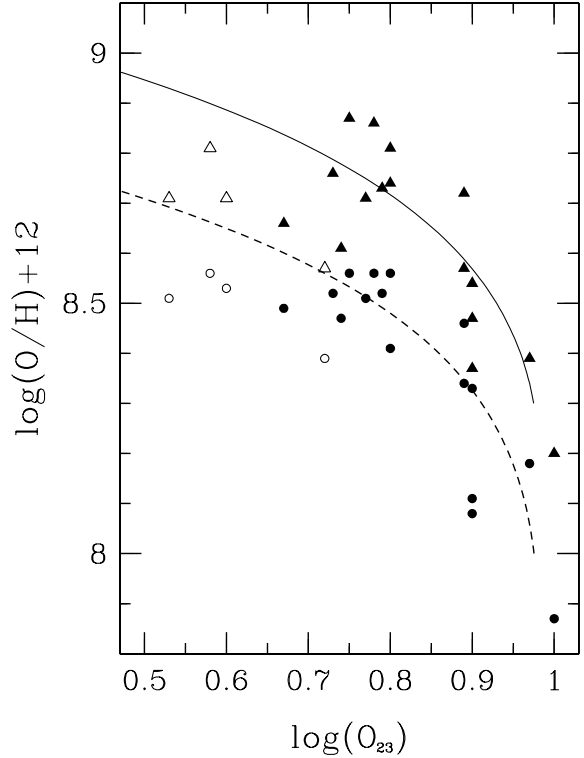}
  \caption{Pagel's $O_{23}$ abundance indicator calibrated 
    using abundances 
    determined with recombination lines --- \ion{O}{2}$_{RL}$ method 
    (triangles and solid line) ---, and abundances 
    determined with collisionally excited lines and $t^2$ = 0.00 
    --- $T$(4363) method (circles and dotted line)---. The
    data is presented in Table 1. The filled symbols correspond to objects
with 0.70 $< P <$ 0.97 and the empty symbols to objects with 0.20 $< P <$ 0.38. 
The lines indicate the calibrations for $P$ = 0.8. To include the  
0.20 $< P <$ 0.38 objects in the  $P$ = 0.8 calibrations we have added to them
0.2 dex in the O/H axis based on the relative difference with $P$ found
by Pilyugin and Thuan (2005).}
  \label{fig:pagel}
\end{figure}

There are three aspects that need to be considered in future work
to have a full calibration of the $O_{23}$ indicator independent 
of temperature variations: a) the calibration of the high metallicity
branch for $O_{23}$ $<$ 0.5, b) the calibration of the $O_{23}$ low
metallicity branch, and c) the variation of the O ionization degree
at a given $O_{23}$ value. We will say a few words about these aspects.

To calibrate the $O_{23}$ indicator for values of log $O_{23}$ $<$ 0.5
in the high metallicity branch we need additional observations of 
\ion{O}{2} lines. In general the higher the metallicity the lower
the degree of ionization and the lower the $N$(O$^{++}$)/$N$(O)
ratio. Therefore when most of the O becomes O$^+$ the \ion{O}{2} 
recombination lines
become very weak and can not be used to derive the O/H values,
consequently we need other temperature indicators to estimate
$T_0$ and  $t^2$. For example the combination of
good Balmer continuum temperatures with temperatures derived from
the $\lambda\lambda 5755, 6584~[\ion{N}{2}]$ lines. Good Balmer
continuum temperatures are difficult to determine due to the 
underlying stellar contribution that contaminates the nebular 
continuum emission.

It is not possible with present day equipment to calibrate 
the $O_{23}$ indicator for the low metallicity branch with 
\ion{O}{2} recombination lines because they become weaker as
$N$(O)/$N$(H) decreases. Fortunately due to the higher $T$(4363)
values in this branch the effect of the temperature variations 
on the $N$(O)/$N$(H) determinations becomes smaller, and we 
expect differences between the $T$(4363) and the $PIM$ methods 
to become smaller than $\sim$ 0.2 dex. Moreover the effect due 
to the temperature structure might be estimated by deriving 
$T_0$ and  $t^2$ from the $T$(4363) and $T$($\ion{He}{1}$) 
temperatures (Peimbert et al. 2000, 2002).

In general for giant very bright \ion{H}{2} regions in the $O_{23}$
low metallicity branch and in the high metallicity branch for
log $O_{23}$ $>$ 0.5 most of the O is twice ionized; but for
old \ion{H}{2} regions and those ionized by a handful of O stars
the fraction of O once ionized becomes important. To test
the effect of the ionization degree in our calibration it is 
necessary to obtain abundances of \ion{H}{2} regions at a given
$O_{23}$ with different O ionization degrees.

\subsection{Calibrations based on photoionization models}

The $PIM$ method is based on photoionization models where O/H is an input of
the models and the observed $O_{23}$ values are adjusted to the predicted
ones. Calibrations based on this method have been presented by many
authors (e. g.: McCall, Rybski, \& Shields 1985; Dopita, \& Evans 1986;
McGaugh 1991; Zaritsky, Kennicutt \& Huchra 1994; Kewley, \& Dopita 2002;
Kobulnicky, \& Kewley 2004). The $PIM$ method depends on
the quality of the models. A good model for the ionizing
cluster should include: an initial mass
function, the time elapsed since the beginning of the star formation, and
a star formation rate; while for the nebula it should include the gaseous density 
distribution.

Photoionization models not yet include all the physical processes needed to
reproduce all the emission line ratios observed in real nebulae. For example 
they do not include the possible presence of stellar winds due to WR stars
nor the possible presence of supernova remnants and related shocks. From a study 
of NGC 604, a giant extragalactic \ion{H}{2} region in M33, Yang et al. (1996)
conclude that the velocity width of the H$\alpha$ line consists of equal
contributions from thermal broadening, stellar winds and SNRs, and gravity.
Even the best
photoionization models, those tailored to fit I~Zw~18, NGC~2363, and NGC~346,
predict $T$(4363) values smaller than observed (Stasi\'nska \& Schaerer 
1999; Luridiana, Peimbert, \& Leitherer 1999, and Rela\~no, Peimbert, 
\& Beckman 2002), probably indicating the need for additional heating sources.
Photoionization models typically predict $t^2 \approx 0.005$, values 
considerably smaller than those derived from observations that are typically 
in the 0.02 $< t^2 < 0.06$
range.

In Figure 2 we present our \ion{O}{2}$_{RL}$ calibration for $P$ = 0.8 and compare it with the
calibrations based on the $PIM$ method by McGaugh (1991) and Kobulnicky \&
Kewley (2004). The agreement between the \ion{O}{2}$_{RL}$ and the $PIM$
calibrations is very good because the $PIM$ calibrations do not fit the
$\lambda$ 4363 line intensity, which depends strongly on $T_0$ and $t^2$, they do fit
the $\lambda\lambda$ 5007 and 3727 line intensities that show a much smaller
dependence on $T_0$ and $t^2$ than the $\lambda$ 4363 line intensity.

\begin{figure}[!t]
  \includegraphics[width=\columnwidth]{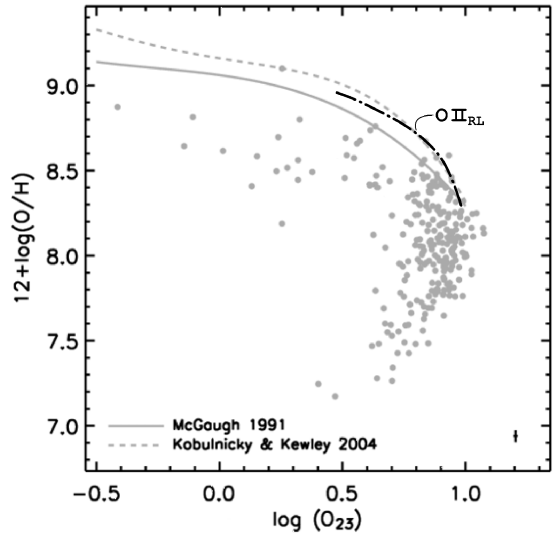}
  \caption{Our \ion{O}{2}$_{RL}$ calibration for $P$ = 0.8 superimposed to a slide
by J. Moustakas presented at the workshop on "Bright Line Abundance Calibrations at Low and High Metallicities" (Minneapolis,  May 2005). The dots represent
O/H ratios determined from the observed $T(4363)$ values under the 
assumption of $t^2$ = 0.00, they are compared with the $PIM$ calibrations by McGaugh (1991) and Kobulnicky \& Kewley (2004). The \ion{O}{2}$_{RL}$ calibration for $P$ = 0.8 is in very good agreement with the $PIM$ calibrations.}
  \label{fig:pagel}
\end{figure}

\subsection{Calibrations based on observations of $O_{23}$ and $T_e(4363)$}

The $T$(4363) method is based on adjusting the observed $O_{23}$ values with
the abundances derived from $T(4363)$ under the assumption that $t^2 = 0.00$. 
The calibrations based on this method depend strongly on the temperature 
structure of the nebulae and underestimate the O/H values by factors of 
about 2 to 3 because $\lambda$ 4363 has a large Boltzmann factor for collisional
excitation that depends strongly on $T_0$ and $t^2$.

The differences between the \ion{O}{2}$_{RL}$ calibration and the observed values
derived with the $T$(4363) calibration for $t^2$ = 0.00
presented in Figure 3 are 
in the 0.2 to 
0.3 dex range and are similar to the differences presented in Table 1,
therefore
we attribute most of the difference between the \ion{O}{2}$_{RL}$
and the $T$(4363) calibrations as being 
due to temperature
variations over the observed volumes. Moreover from the similarity
shown in Figure 2 between the $PIM$ calibration and the \ion{O}{2}$_{RL}$
calibration for $P$ = 0.8 we also conclude that the main difference between 
the $PIM$
calibration and the $T$(4363) calibration is due to temperature variations
over the observed volume.

In their recent calibration of the $T$(4363) method for different 
$P$ values, Pilyugin and Thuan (2005) adopted the temperature for 
the once ionized O region given by $T$(O$^+$) = 0.7$\times$$T$(4363) + 3000 K (Garnett 1992) to derive the $N$(O$^+$)/$N$(H$^+$)
ratios. Therefore the $T$(O$^+$) and the $N$(O$^+$)/$N$(H$^+$) values so 
derived also depend 
on the possible presence of temperature variations in the O$^{++}$ regions.

\begin{figure}[!t]
  \includegraphics[height=\columnwidth,angle=270]{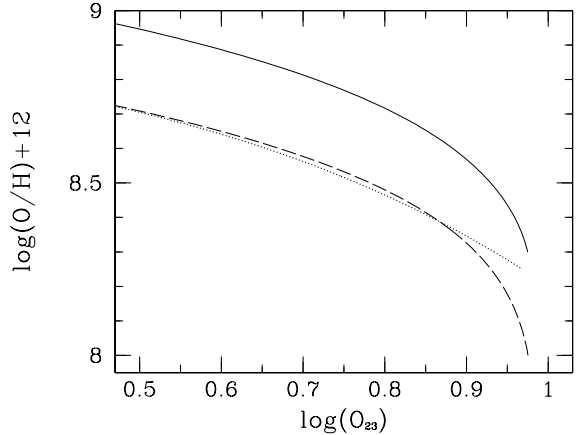}
  \caption{Comparison of three absolute calibrations for $P$ = 0.8.
The solid line represents our \ion{O}{2}$_{RL}$ calibration,
the dashed line represents our $T$(4363) calibration, and the dotted line
the Pilyugin and Thuan (2005) calibration. Note the excellent agreement
between the $T$(4363) calibrations, this agreement implies that the sole
difference between our \ion{O}{2}$_{RL}$ calibration and the
$T$(4363) calibration by Pilyugin and Thuan is due to temperature 
variations inside the observed objects.} 
  \label{fig:pagel}
\end{figure}

\begin{table}[!t]\centering
  \setlength{\tabnotewidth}{\columnwidth}
  \tablecols{5}
  \caption{Average O/H values for 14 disk galaxies}
\label{tab:Average}
\begin{tabular}{lcccc}
    \hline
    \hline
    {\small Method} & {\small $\left<{\rm O/H}\right>$} 
& {\small $\left<{\rm O/H}\right>$} & {\small $\left<{\rm Gradient}\right>$}
& {\small {Cal.} } \\
 & {\small $\rho$ = 0} & {\small $\rho=0.4\rho_{25}$} & {\small dex$\rho^{-1}_{25}$}\\
    \hline
$PIM$        & 9.14 & 8.86 & -0.48 & 1 \\
$T$(4363)    & 8.58 & 8.38 & -0.28 & 2 \\
    \hline
    \hline
    \tabnotetext{}{Results by Moustakas \& Kennicutt (2006), O/H values
    given in 12 + log O/H.}
    \tabnotetext{}{1- McGaugh(1991); 2- Pilyugin,  \& Thuan (2005).}
\end{tabular}
\end{table}

In Table 2 we present the averaged O/H determinations by Moustakas \&
Kennicutt (2006) based on $O_{23}$ observations of 234 \ion{H}{2} 
regions using the $PIM$ calibration
by McGaugh (1991) and the $T$(4363) calibration by Thuan and Pilyugin
(2005). The galactocentric distance is given by $\rho$, 
the O/H value at $\rho$ = 0 corresponds to
the extrapolation to the galactic center, and $\rho_{25}$ is
the radius of the semi-major axis at the $B_{25}$ mag arc sec$^{-2}$
isophote. For this sample the abundance controversy amounts to 0.56 dex
for $\rho$ = 0, and to 0.48 dex for $\rho=0.4\rho_{25}$. The increase 
in the O/H difference with metallicity between both calibrations,
as well as the steeper abundance gradient for the $PIM$ calibration 
relative to the $T$(4363) calibration are consistent with the idea that
temperature variations are mainly responsible for these differences.
The larger differences at larger O/H values are expected due to the higher 
sensitivity of O/H on $T_0$ and $t^2$ as $T$(4363) becomes smaller, in other words
the larger differences are due to the Boltzmann factor for collisional excitation of the $\lambda$
4363 line that becomes larger at smaller $T$(4363) values.

\section{Galactic \ion{H}{2} regions and the solar abundances}
\label{sec:Gal}

In addition to the evidence presented in section 2.1 in favor
of large $t^2$ values, and consequently in favor of the \ion{O}{2}$_{RL}$ 
method, there is another independent
test that can be used to discriminate between the $T$(4363) method and 
the \ion{O}{2}$_{RL}$ method that consists
in the comparison of stellar and  \ion{H}{2} region abundances of the 
solar vicinity. To carry out this comparison we
have added 0.08 dex to all the gaseous O/H  determinations to take into 
account the estimated fraction of O tied up in dust grains in \ion{H}{2} regions 
(see Esteban et al. 1998).

Esteban et al. (2005) determined that 12 + log (O/H) = 8.77 for 
the ISM of the solar vicinity based on the O/H galactic gradient 
derived from the \ion{O}{2}$_{RL}$ method. Alternatively from the 
solar ratio by 
Asplund, Grevesse, \& Sauval (2005), 
that amounts to 12 + log(O/H) = 8.66, and taking into account
the increase of the O/H ratio due to galactic chemical evolution since
the Sun was formed, that according to state of the art chemical
evolution models of the Galaxy amounts to 0.13~dex (e.g. Carigi et
al. 2005), we obtain an O/H value of 8.79~dex, in excellent agreement 
with the value based on 
the \ion{O}{2}$_{RL}$ method. In this comparison we are assuming that 
the solar abundances 
are representative of the abundances of the solar vicinity ISM when
the Sun was formed.

There are two other determinations of the present
O/H value in the ISM that can be made from 
observations of F and G stars of the solar vicinity. According to 
Allende-Prieto et al. (2004) the Sun appears deficient by roughly 
0.1 dex in O, Si, Ca, Sc, Ti, Y, Ce, Nd, and Eu, compared with its 
immediate neighbors with similar iron abundances, by adding this 
0.1 dex difference to the 
solar value by Asplund et al. (2005) we obtain a lower limit of 
12 + log O/H = 8.76 for the local interstellar medium.
A similar result is obtained from the data
by Bensby \& Feltzing (2005) who obtain for the six most O-rich 
thin-disk F and G dwarfs of the solar vicinity an average  
[O/H] = 0.16; by adopting their value as representative of the 
present day ISM of the solar vicinity we find 12 + log O/H = 8.82. 
Both results are in excellent agreement with the O/H value derived 
from the \ion{O}{2}$_{RL}$ method.

On the other hand, based on the $T$(4363) method with $t^2$ = 0.00
Deharveng et al. (2000) 
and Pilyugin, Ferrini, \& Shkvarun (2003) obtain 12 + log O/H values of 
8.61 and 8.60 respectively for the solar vicinity, values about 0.2 dex
smaller than the stellar predictions and the value derived from the
\ion{O}{2}$_{RL}$ method.

\section{Calibration of other metallicity indicators}
\label{sec:other}

There are other O/H indicators that have been proposed in
the literature: $O_3N_2\equiv 
I([$\ion{O}{3}$]\lambda 5007/I([$\ion{N}{2}$]\lambda 6584)$, presented 
by Alloin et al. (1979), $N_2\equiv I([$\ion{N}{2}$]\lambda 6584)/I({\rm H}(\alpha)$, 
presented by Storchi-Bergmann et al. (1994), 
$S_{23}\equiv I([$\ion{S}{2}$]\lambda\lambda 6717,6731 + [$\ion{S}{3}$]\lambda 9069)/ I({\rm H}\alpha)$, 
presented by 
V\'\i{}lchez, \& Esteban (1996), 
$Ar_3O_3\equiv I([$\ion{Ar}{3}$]\lambda 7135/I([$\ion{O}{3}$]\lambda 5007)$, 
and $S_3O_3\equiv I([$\ion{S}{3}$]\lambda 9069/I([$\ion{O}{3}$]\lambda 5007)$, 
these two proposed by Stasi\'nska 
(2006). Since these indicators are calibrated with models
that fit the nebular O/H lines,
or with O/H determinations based on $T$(4363) observations,
the absolute calibration shift derived for the $O_{23}$ calibration
based on the \ion{O}{2}$_{RL}$ method
also applies to them. Therefore the O/H values derived from them have
to be increased, as is the case for the $O_{23}$ indicator, if they 
are calibrated with the $T(4363)$ method.

In addition to the absolute calibration shift the $N_2$
indicator has other problems: a) it shows a larger dispersion than 
the other indicators making it less reliable (e. g. Stasi\'nska 2006), 
b) according to
Stasi\'nska (2006) and Moustakas and Kennicutt (2006) the possible 
contribution to the
$I([$\ion{N}{2}$]\lambda 6584)/ I({\rm H}\alpha)$ ratio by the extended 
low density interstellar medium, that is expected to be more important 
for galaxies farther away, also might produce a bias in the calibration 
of the $N_2$ indicator,
c) the indicators based on O, S, and Ar have the advantage that these are 
primary elements formed by massive stars, therefore their relative 
abundance ratios are almost constant during the chemical
evolution of galaxies, this is not the case for N that is produced 
by two types of stars, massive (that end their lives as supernovae)  
and low and intermediate mass stars (that end their lives as white dwarfs), 
and in two different ways from C and O produced by their own star (primary
origin), or using C and O of the progenitor cloud where the star formed 
(secondary origin); moreover, other effects, as stellar rotation 
(Meynet \& Maeder 2002) and the treatment of hot bottom burning, 
cause substantial differences in the computed N yields
(see the compilation by Gavil\'an, Moll\'a, \& Buell 2006).

If the $N_2$ or the $O_3N_2$ indicators based on observations of nearby
galaxies, are used for objects at large distances, considering that we 
are comparing two sets of different ages, the variation in 
N/O as a function of time predicted by the chemical evolution models 
for galaxies of different masses and different star formation histories 
has to be considered.

\section{Conclusions}
\label{sec:conc}

The \ion{O}{2}$_{RL}$ method supports the suggestion that the controversy
produced by the relatively high O/H values predicted
by the $PIM$ calibrations and the relatively low O/H
values predicted by the  $T(4363)$
calibrations are mainly due to temperature variations.

The best way to calibrate the $O_{23}$ indicator is to use the
 \ion{O}{2}$_{RL}$ method to obtain the O/H values because it is
independent of the temperature structure.

The use of $T$(4363) values to derive O/H, under the assumption of
constant temperature, provides a lower limit to the O/H abundance ratios.

Since the nebular lines are less sensitive to $T_0$ and $t^2$ than the auroral lines, the model calibrations that adjust the nebular lines are closer 
to the \ion{O}{2}$_{RL}$ calibration than those derived using the 
observed $T$(4363) values.

For a given object the \ion{O}{2} recombination lines provide gaseous 
abundances that are 
about 0.2 to 0.3 dex higher than those derived from collisionally 
excited lines and $T$(4363) under the assumption that $t^2$ = 0.00.

By using the  \ion{O}{2} recombination lines to derive O/H abundances in Galactic
\ion{H}{2} regions together with state of the art Galactic chemical
evolution models we obtain an excellent agreement with the O/H
solar value derived by Asplund et al.(2005). We also find
an excellent agreement between the \ion{H}{2} region O/H abundances
derived from recombination lines and those derived from young G dwarfs, 
this comparison is
independent of Galactic chemical evolution effects since
both types of objects correspond to the present abundances
in the ISM of the solar vicinity.

With the \ion{O}{2}$_{RL}$ calibration and observations of emission line
galaxies at different $z$ values it will be possible to study
the chemical evolution of the universe as a whole. All the other 
O/H indicators, like those known as $O_3N_2$, 
$N_2H\alpha$, $S_{23}$, $Ar_3O_3$, and $S_3O_3$, depend on an 
absolute calibration and we recommend to calibrate them using 
the \ion{O}{2}$_{RL}$ method.

Models of the chemical evolution of N/O versus O/H as a function
of time for different types of galaxies are required to calibrate
the $N_2$ and $O_3N_2$ indicators. 

We need more high resolution observations of \ion{O}{2} recombination
lines in the local universe to refine and extend the \ion{O}{2}$_{RL}$
calibration. We need objects with 
log $O_{23}$ smaller than 0.5 in the high metallicity branch.
We also need  objects of different $P$ values 
at a given $O_{23}$ value in the high metallicity branch.

\acknowledgments
It is a pleasure to acknowledge fruitful discussions with 
Evan Skillman, Bob Kennicutt, Henry Lee, and Valentina Luridiana. MP is
grateful to our Florida colleagues for their warm hospitality
during this meeting.

\end{document}